\documentstyle[aps,12pt,preprint,prb,tabularx]{revtex}

\begin{document}
\title{Complexation of oppositely charged polyelectrolytes: effect of ion
pair formation.}
\author{Alexander Kudlay\thanks{E-mail: kudlay@northwestern.edu}, Alexander V. Ermoshkin,\thanks{%
E-mail: ermosh@email.unc.edu. Present address: Department of
Chemistry, University of North Carolina, Chapel Hill, North
Carolina, 27599} and Monica Olvera de la Cruz\thanks{E-mail:
m-olvera@northwestern.edu}}
\address{Department of Materials Science and Engineering\\
Northwestern University, Evanston, Illinois 60208}
\maketitle

\begin{abstract}
Complexation in symmetric solutions of oppositely charged polyelectrolytes
is studied theoretically. We include polyion crosslinking due to formation
of thermoreversible ionic pairs. The electrostatic free energy is calculated
within the Random Phase Approximation taking into account the structure of
thermoreversible polyion clusters. The degree of ion association is obtained
self-consistently from a modified law of mass action, which includes
long-range electrostatic contributions. We analyze the relative importance
of the three complexation driving forces: long-range electrostatics, ion
association and van der Waals attraction. The conditions on the parameters
of the system that ensure stability of the complex with addition of salt are
determined.
\end{abstract}

\newpage

\section{Introduction}

The complexation of oppositely charged polyelectrolytes in different ionic
conditions is an interesting problem of biological relevance.\cite%
{Blomfieldreviw,Raspaud98,Levin02,k91} Moreover, the association of two
oppositely charged linear chains has implications in the design of new
materials with unique properties such as multilayer polyelectrolytes\cite%
{de97,m04}, and carrier gels for drug delivery\cite{drug}. Industrial
applications include uses as coatings, flocculants and absorbents.\cite%
{flocc,abs}

Charged chains can associate via many mechanisms given the large number of
length scales involved in these mixtures. The properties of the complexes
are strongly dependent on physical parameters such as salt and monomer
concentrations, and on the chemical configurations of the chains such as
their charge densities, persistence lengths, degrees of polymerization, and
nature of the interactions between the charge groups along the backbone.\cite%
{k94,ger,d00,ts94,p97}

Complexes formed by chains with low charge densities have been analyzed
using linearized models.\cite{be90,c01,c02,mine,cs04} These models are
applicable to describe solutions with strongly hydrated charge groups of
positive and negative charges, which interact weakly with each other. In
these studies ion condensation effects\cite{m69,All} can be neglected if the
linear charge density is sufficiently low. The complexation in these systems
occurs via collective charge fluctuations.\cite{be90,c01,c02,mine} With
addition of salt the complexes dissolve due to screening of electrostatic
interactions.

In certain polymer mixtures, even when the charge density is low, non-linear
effects are important if the charge groups are strongly interacting. In
these mixtures the oppositely charged groups, when placed at short
separation distances, can be locally dehydrated, and act as localized
short-range crosslinks between oppositely charged groups along the chains.
The formation of these links generates a thermoreversibly crosslinked
solution\cite{ee99,we,dd04} of oppositely charged chains. The number of
crosslinks formed in equilibrium in such thermoreversibly associating chains
is given by the law of mass action with an effective association constant.
In uncharged systems the association constant depends exponentially on the
strength of the short range attraction between the reactive groups. For
charged reactive groups, however, there is an additional term in the
association constant due to the electrostatic contribution to the free
energy resulting from the collective charge fluctuations. Since association
occurs between the ions belonging to oppositely charged chains, this
additional electrostatic term always increases the rate of association. The
importance of non-linear association of charges has been recently recognized
in polyelectrolyte adsorption\cite{m01,g98} and multilayer formation\cite%
{m02}.

In this work the electrostatic interactions are described using a two-fold
approach. The strongly non-linear short-range interactions between
oppositely charged groups are accounted for by including strong correlations
between the chains. The long-range electrostatic interactions (which are
weak for weakly charged polyelectrolytes) are accounted for in a linearized
way by computing the fluctuations of this correlated solution of crosslinked
charged chains using a generalized Debye-H\"{u}ckel approach (Random Phase
Approximation).\cite{be88,jl90} The electrostatic free energy depends on\
the structure of charged polymer clusters, while the cluster distribution
depends (through the modified law of mass action) on the electrostatic free
energy. Therefore, we determine here the number of formed crosslinks by
evaluating self-consistently the electrostatic contribution from the
collective charge fluctuations of a crosslinked system of charged chains.

The degree of hydrophobicity of the chain backbone modifies the
thermodynamics of the solution.\cite{mine} We consider non-selective
solvents, where the degree of compatibility is the same for both the
positively and negatively charged chains. We investigate how the degree of
hydrophobicity influences the properties of complexes formed by
electrostatic interactions.

The paper is organized as follows. In Section \ref{model} we describe the
model and approximations used, and derive the free energy of the solution;
the details of the derivation of the correlation function of the crosslinked
system required to determine the electrostatic contribution to the free
energy is given in Appendix \ref{app}. In Section \ref{express} we discuss
how different system parameters influence the properties of the formed
complex and its response to addition of salt. The conclusions are given in
Section \ref{conclusions}.

\section{The free energy of the semi-dilute solution}

\label{model}

\subsection{Model}

In this section we calculate the free energy of a homogeneous semidilute
solution of oppositely charged polyelectrolytes. For simplicity in this work
we consider only the absolutely symmetric case. Positively and negatively
charged chains are present in the solution in equal concentrations, and have
the\ same physical properties except \ the sign of the charge. The chains
have degree of polymerization $N=N_{1}=N_{2}$. The fraction of charged
monomers on both types of chains is equal to $f$. We consider only weakly
charged chains for which $f$ is small enough, so that electrostatic energy
of adjacent along the chain charges is smaller than the thermal energy.\cite%
{kh82,d95} The number concentrations of positively and negatively charged
monomeric units are $\rho _{1}=\rho _{2}$, so that the total concentration
of monomeric units in the solution is $\rho =2\rho _{1}/f$. Each charged
monomeric unit releases a monovalent counterion. A 1:1 low molecular salt
can also be present in the solution with concentration of positively and
negatively charged ions given by $\rho _{s+}$ and $\rho _{s-}$,
respectively, and the total concentration of salt ions $\rho _{s}=2\rho
_{s+}=2\rho _{s-}$. Since the effect of counterions is equivalent to
addition of salt we include the counterions in the salt.

We describe strong electrostatic interactions between oppositely charged
monomeric units as thermoreversible bond (crosslink) formation. This strong
non-linear interaction should be treated differently from the long-range
electrostatic part which is treated within the Random Phase Approximation
(RPA).\cite{be88,jl90} Another reason why formation of ionic pairs should be
considered in addition to Coulomb interactions is because formation of ionic
pairs can proceed with the rearrangement of the solvation shell of charges
on polymer. In this case the effective dielectric constant between charges
in a pair can differ significantly from the bulk solvent dielectric
constant. Thus, the bond energy in a pair can be quite high. A natural model
to describe such ion-pairing is reversible association between charges of
opposite signs. For simplicity we assume that only pairs can be formed, with
the absolute value of the reduced bond energy $\varepsilon =\left\vert
E\right\vert /kT$ ($k$ is the Boltzmann constant, $T$ the thermodynamic
temperature), which gives rise to association constant $\omega
=e^{\varepsilon }$. Note that bond formation is different from short-range
van der Waals attraction in that it has the saturation property, that is,
once a bond between two given ions is formed they do not interact with any
other ions.

We write the free energy of the solution of associating polyelectrolytes in
the following form%
\begin{equation}
F=F_{ref}+F_{RPA}  \label{f11}
\end{equation}%
where the first term is the free energy of the reference neutral system (but
with short-range interactions) and the second term is the contribution of
electrostatics. The electrostatic part $F_{RPA}$ is calculated within the
RPA, which is a linear theory equivalent to the Debye-H\"{u}ckel
approximation (i.e., to the linearized Poisson-Boltzmann equation). For the
reference free energy we use the Flory-Huggins mean-field approximation
\begin{equation}
F_{ref}=F_{id}+F_{FH}  \label{fref}
\end{equation}%
which includes the ideal entropic and enthalpic terms. In our associating
system the ideal term is the free energy of ideal gas of all possible
clusters $\{C\}$ with appropriate statistical weights $\omega (C)$%
\begin{equation}
\frac{F_{id}}{kTV}=\sum_{\{C\}}\rho (C)\ln \frac{\rho (C)}{e\omega (C)}+\rho
_{s-}\ln \frac{\rho _{s-}}{e}+\rho _{s+}\ln \frac{\rho _{s+}}{e}
\end{equation}%
as well the entropy of the ideal gas of salt ions (counterions are also
included here). As has been shown in the refs \CITE{ee99,we} the equilibrium
concentrations $\rho (C)$ can be obtained using a diagrammatic technique and
the free energy of associating chains can be written as%
\begin{equation}
\sum_{\{C\}}\rho (C)\ln \frac{\rho (C)}{e\omega (C)}=\frac{\rho }{N}\ln \rho
+\rho f\left[ (1-\Gamma )\ln (1-\Gamma )+\Gamma \ln \Gamma \right] -\frac{%
\rho f\Gamma }{2}\ln \left[ \frac{\rho f\Gamma }{2e}ve^{\varepsilon }\right]
\label{ass}
\end{equation}%
Here conversion $\Gamma $ is the fraction of polymeric ions in pairs (see
Appendix \ref{app}), which is to be found from subsequent minimization of
the total free energy of the solution. The volume of the monomeric unit $v$
(which we for simplicity assume to be equal to $v=b^{3}$) is used to
approximate the internal partition function of the crosslink $%
Z_{cross}=ve^{\varepsilon }$. Alternative combinatorial derivation of (\ref%
{ass}) can be performed along the lines of ref \CITE{sem}.

We assume that the polycation and polyanion backbones have identical short
range interaction with solvent. The interaction free energy in (\ref{fref})
is assumed to be given by the Flory-Huggins form
\begin{equation}
\frac{F_{FH}}{kTV/b^{3}}=(1-\phi -\phi _{s})\ln (1-\phi -\phi _{s})+\chi
\phi (1-\phi )  \label{ffh}
\end{equation}%
where $V$ is the volume of the system, $\phi =\rho b^{3}$ is the total
polymer volume fraction, and $\phi _{s}=b^{3}\rho _{s}=2b^{3}\rho _{s-}$ is
the total volume fraction of salt ions (and also counterions). The first
term in (\ref{ffh}) stems from hardcore repulsion; the second one from
short-range attraction, whose strength is characterized by the parameter $%
\chi $. We will analyze both the cases of good and marginal to bad solvent.
Note that, in contrast to previous works,\cite{c01,kh92} interactions of
backbones with solvent favor complexation under bad solvent conditions.

Adding up the two contributions, the free energy $F_{ref}$ \ of the
reference neutral system reads%
\begin{eqnarray}
\frac{F_{ref}}{kTV/v} &=&\frac{\phi }{N}\ln \phi +\phi _{s}\ln \phi
_{s}+\phi f\left[ (1-\Gamma )\ln (1-\Gamma )+\Gamma \ln \Gamma \right] -%
\frac{\phi f\Gamma }{2}\ln \left[ \frac{\phi f\Gamma }{2e}e^{\varepsilon }%
\right] +  \nonumber \\
&&+(1-\phi -\phi _{s})\ln (1-\phi -\phi _{s})+\chi \phi (1-\phi )
\label{fref2}
\end{eqnarray}

\subsection{Electrostatic free energy: Random Phase Approximation}

Due to electroneutrality the electrostatic contribution $F_{RPA}$ in the
total free energy (\ref{f11}) is due to fluctuations of charge
concentration, which is calculated within the Random Phase Approximation
(see ref \CITE{we} for details):%
\begin{equation}
\frac{F_{RPA}}{kT}=\frac{V}{2}\int \frac{d^{3}q}{(2\pi )^{3}}\left[ \ln
\left( \det \left( {\bf I}+{\bf G}(q){\bf U}(q)\right) \right) -\sum_{i}\rho
_{i}U_{ii}(q)\right]   \label{frpa}
\end{equation}%
where ${\bf I=}\left\vert \left\vert {\bf \delta }_{ij}\right\vert
\right\vert $ is the unitary matrix, ${\bf G}(q)$ is the correlation
function matrix of the reference neutral system and ${\bf U}(q)$ is the
matrix of Coulomb interactions. The sum runs over all charged components of
the system (co-ions and salt ions). The last term in (\ref{frpa}) is the
self-energy of pointlike charges. The correlation function matrix ${\bf G}(q)
$ can be in turn obtained within the RPA as\cite{iya,deGennes,benoit}%
\begin{equation}
{\bf G}^{-1}(q)={\bf g}^{-1}(q)+{\bf c}(q)
\end{equation}%
where ${\bf g}(q)$ is the structure correlation matrix. It characterizes
correlations of density due to existence of different clusters in the
system, but does not include interactions. The matrix ${\bf g}(q)$ for our
symmetric system has the form

\begin{equation}
g_{ij}=\left(
\begin{array}{cccc}
g_{11} & g_{12} & 0 & 0 \\
g_{12} & g_{11} & 0 & 0 \\
0 & 0 & \rho _{s-} & 0 \\
0 & 0 & 0 & \rho _{s+}%
\end{array}%
\right)   \label{gg}
\end{equation}%
The polymeric correlation function $g_{11}$ and $g_{12}$ are calculated in
Appendix \ref{app} using a diagrammatic approach. The interaction matrix $%
{\bf c}(q)$ describes short-range interactions (free energy $F_{FH}$) and
its components are given by%
\begin{equation}
c_{ij}=\frac{1}{\Phi }s_{i}s_{j}-2\chi p_{i}p_{j}  \label{c1}
\end{equation}%
where we introduced the volume fraction of solvent $\Phi =1-\phi -\phi _{s}$
and two auxiliary vectors
\begin{eqnarray}
s_{i} &=&\{1,1,1,1\} \\
p_{i} &=&\{1,1,0,0\}  \label{c3}
\end{eqnarray}%
Using the vector of valencies $e_{i}$, the Coulomb interaction matrix can be
written in the following form
\begin{eqnarray}
U_{ij}(q) &=&e_{i}e_{j}U(q) \\
e_{i} &=&\{1,-1,1,-1\}
\end{eqnarray}%
which allows us to simplify the expression under the logarithm in (\ref{frpa}%
)%
\begin{equation}
\det \left( {\bf I}+{\bf G}(q){\bf U}(q)\right)
=1+U(q)\sum_{i.j}G_{ij}(q)e_{i}e_{j}  \label{det}
\end{equation}%
Using formulae (\ref{c1}) and (\ref{c3}) for the correlation functions $%
g_{ij}$ and $c_{ij}$ we now can obtain $G_{ij}$ in accordance with (\ref{gg}%
). Substituting the result into (\ref{det}) we finally obtain%
\begin{equation}
\det \left( {\bf I}+{\bf G}(q){\bf U}(q)\right) =1+U(q)\left\{
2g_{11}-2g_{12}+2\rho _{s-}\right\}   \label{det2}
\end{equation}%
It is remarkable that this result is the same as if we had neglected the
short-range interactions (the matrix $c_{ij}$), that is, if we had put $%
G_{ij}=g_{ij}$ in determinant (\ref{det}). Note, however, that this simple
result holds only for\ our case of a symmetric system (described by matrix $%
g_{ij}$) and for symmetric long-range (matrix ${\bf U}(q)$) and short-range
(matrix ${\bf c}$) interactions.

The structure correlation functions $g_{ij}$ are calculated in Appendix \ref%
{app} (see (\ref{gf11}--\ref{gf12})). We reproduce them here for convenience%
\begin{eqnarray}
g_{11}(q) &=&\rho _{1}g(q)\frac{1+\left( \Gamma ^{\prime }\right) ^{2}h(q)}{%
1-\left[ \Gamma ^{\prime }h(q)\right] ^{2}}  \label{pc1} \\
g_{12}(q) &=&\rho _{1}g(q)\frac{\Gamma ^{\prime }g(q)}{1-\left[ \Gamma
^{\prime }h(q)\right] ^{2}}  \label{pc2}
\end{eqnarray}%
The functions $g(q)$ and $h(q)$ are defined by (\ref{sg}) and (\ref{hf}) in
Appendix \ref{app}. The effective conversion $\Gamma ^{\prime }$ is defined
as $\Gamma e^{-q^{2}b^{2}/6}$ in (\ref{ec}), with the bare conversion $%
\Gamma $ defined as the fraction of charged monomers participating in
crosslinks (see eq \ref{dg}). It is important to note that the correlation
functions (\ref{pc1}--\ref{pc2}) are calculated for an ideal
thermoreversibly associating system, in which no other interactions except
crosslinking are present (in our case no electrostatic and hydrophobic
interactions).\cite{we}

Substituting these expressions into (\ref{det2}) we can rewrite the free
energy (\ref{frpa}) as%
\begin{equation}
\frac{F_{RPA}}{kT}=\frac{V}{2}\int \frac{d^{3}q}{(2\pi )^{3}}\left[ \ln
\left( 1+U(q)\left\{ \rho fg(q)\frac{1-\Gamma ^{\prime }}{1+\Gamma ^{\prime
}h(q)}+\rho _{s}\right\} \right) -\sum_{i}\rho _{i}U_{ii}(q)\right]
\label{frpa2}
\end{equation}%
where $\rho f=2\rho _{1}=2\rho _{2}$, and $\rho _{s}=2\rho _{s-}=2\rho _{s+}$%
.

To be able to evaluate\ $F_{RPA}$ we need to specify a suitable form of the
interaction potential $U(q)$. For bare Coulomb interaction we have%
\begin{eqnarray}
\frac{U_{C}(r)}{kT} &=&\frac{q_{e}^{2}}{\epsilon kT}\frac{1}{r}=\frac{l}{r}
\label{co1} \\
\frac{U_{C}(q)}{kT} &=&\int d^{3}r~e^{i{\bf qr}}U_{C}(r)=\frac{4\pi l}{q^{2}}
\label{co2}
\end{eqnarray}%
Here we introduced the Bjerrum length $l=q_{e}^{2}/(\epsilon kT)$, where $%
q_{e}$ is the electron charge and $\epsilon $ the dielectric constant of the
solvent. In order to take into account the influence of the hardcore of the
ions on the electrostatic contribution $F_{RPA}$ we use a modified Coulomb
potential given by%
\begin{eqnarray}
\frac{U(r)}{kT} &=&\frac{l}{r}\left( 1-e^{-r/b}\right)   \label{mo1} \\
\frac{U(q)}{kT} &=&\frac{4\pi l}{q^{2}(1+q^{2}b^{2})}  \label{mo2}
\end{eqnarray}%
where the bond length $b$ is for simplicity taken to be the size of the
ions. At large distances ($r\gg b$) the modified potential becomes the pure
Coulomb potential (\ref{co1}--\ref{co2}). However, at $r=0$ the modified
potential attains a finite value, while the original Coulomb potential
diverges. Thus we phenomenologically include the impenetrability of the ions
within the RPA formalism, which is originally formulated for pointlike ions.
The RPA with the modified potential (\ref{mo1}--\ref{mo2})\ has been shown
to successfully describe the phase diagrams of polyelectrolytes\cite{emac}
and of the low-molecular system of charged dumbbells.\cite{rpm} Furthermore,
this potential has been successfully used in the liquid state approaches.%
\cite{gm03}

Substituting the modified potential (\ref{mo2}) into (\ref{frpa2}) we obtain
the final expression for the electrostatic free energy%
\begin{equation}
\frac{F_{RPA}}{kT}=\frac{V}{2}\int \frac{d^{3}q}{(2\pi )^{3}}\left[ \ln
\left( 1+\frac{4\pi l}{q^{2}(1+q^{2}b^{2})}\left\{ \rho fg(q)\frac{1-\Gamma
^{\prime }}{1+\Gamma ^{\prime }h(q)}+\rho _{s}\right\} \right) -\frac{4\pi l%
}{q^{2}(1+q^{2}b^{2})}(\rho f+\rho _{s})\right]  \label{frpa3}
\end{equation}%
Let us make several comments on the structure of $F_{RPA}$. The polymer
structure correlation functions (\ref{pc1}--\ref{mo2}) diverge at the
gelation condition $\Gamma (Nf-1)=1$, which is simply due to the fact that
gelation corresponds to the formation of \ an \ infinite cluster. It is
remarkable that the electrostatic free energy (\ref{frpa3}) has no
corresponding singularity at the gelation structural transition. This is due
to the charge symmetry of the considered system. Indeed since in our case
association is possible only between oppositely charged chains (which carry
the same amount of charge) the infinite cluster is by construction neutral,
therefore does not contribute to $F_{RPA}$. (It can be shown that for any
asymmetric system (asymmetry of $N$, $\rho $ or $f$) the infinite cluster is
charged and, accordingly, expression for $F_{RPA}$ has a singularity at and
beyond the gelation transition. However, this unphysical singularity is an
artifact of our simplified description of gel structure.)

Let us look at the limiting cases of eq \ref{frpa3}. By putting  $\Gamma
^{\prime }=0$ we regain \ the well-known expression for free unassociated
chains, and if we put $\Gamma ^{\prime }=1$, the polyelectrolyte chains do
not contribute to $F_{RPA}$. However, since $\Gamma ^{\prime }$ is only the
effective conversion:$\ \Gamma ^{\prime }=\Gamma e^{-q^{2}b^{2}/6}$ the
equality $\Gamma ^{\prime }=1$ is possible only when $\Gamma =1$ and $q=0$.
On all finite lengthscales ($q\neq 0$) charges in crosslinks still
contribute to the electrostatic free energy $F_{RPA}$. This is natural,
since in our model of crosslinking the two opposite charges do not
annihilate, rather they are considered as separate charges with Gaussian
correlations between them, which leads to the emergence of effective
conversion $\Gamma ^{\prime }$, instead of bare conversion $\Gamma $.
Because the charges do not annihilate when the ionic pairs are formed, we
subtract the self-energy of all ions present in the system (last term in (%
\ref{frpa3})), regardless of whether they are free or form crosslinks.

The chain correlation functions $g(q)$ and $h(q)$ are defined by (\ref{sg})
and (\ref{hf}). The function $g(q)$ can be easily calculated in the
continuous limit, the result being the well-known Debye structure function.
However, we also need the correct limit of pointlike ions at $q\rightarrow
\infty $ in (\ref{frpa3}), since we subtract the self-energy of all ions
Therefore, we choose a simple interpolation form for the chain structure
function $g(q)$%
\begin{eqnarray}
g(q) &=&1+\frac{Nf}{1+q^{2}b^{2}N/12} \\
h(q) &=&g(q)-1
\end{eqnarray}%
which gives the correct limit at $q=0$, has the scaling of a Gaussian chain
at $N^{-1/2}\ll qb\ll f^{1/2}$, and reproduces pointlike ions at $qb\gg
f^{1/2}$.

\subsection{Minimization of the free energy\label{secmin}}

The total free energy (\ref{f11}) is given by the sum of $F_{ref}$ in (\ref%
{fref2}) and $F_{RPA}$ in (\ref{frpa3}). However, this is only the virtual
free energy of a system with a given value of conversion $\Gamma $. To
obtain the equilibrium free energy $F(\Gamma _{eq})$ we need to obtain the
equilibrium conversion $\Gamma _{eq}$, by the minimization of $F(\Gamma )$:%
\begin{equation}
\frac{\partial F(\Gamma )}{\partial \Gamma }=0
\end{equation}%
Using (\ref{fref2}) and (\ref{frpa3}) we obtain the following equation for $%
\Gamma $%
\begin{equation}
\frac{\Gamma }{\left( 1-\Gamma \right) ^{2}}=\frac{\phi f}{2}\exp \left[
\varepsilon +\mu _{RPA}(\Gamma )\right]   \label{mal}
\end{equation}%
This equation has the general structure of the law of mass action. A
noteworthy feature of (\ref{mal}), however, is that the energy gained from
formation of a crosslink consists of the bonding energy $\varepsilon $ and
the energy gain $\mu _{RPA}$ resulting from the long-range electrostatic
attraction of the polymer chains described by $F_{RPA}$:%
\begin{eqnarray}
\mu _{RPA}(\Gamma ) &=&\int \frac{d^{3}q}{(2\pi )^{3}}\frac{U(q)}{1+U(q)K(q)}%
\frac{g^{2}(q)e^{-q^{2}b^{2}/6}}{\left[ 1+\Gamma ^{\prime }h(q)\right] ^{2}}
\label{mg} \\
U(q) &=&\frac{4\pi l}{q^{2}}\frac{1}{1+q^{2}a^{2}} \\
K(q) &=&\rho fg(q)\frac{1-\Gamma ^{\prime }}{1+\Gamma ^{\prime }h(q)}+\rho
_{s}
\end{eqnarray}%
Of course the energy $\mu _{RPA}$ depends on the thermodynamic state of the
solution, and thus on $\Gamma $, therefore equations (\ref{mal}) and (\ref%
{mg}) are to be solved simultaneously to obtain $\Gamma (\phi )$. Since in
our model crosslinking takes place only between oppositely charged chains $%
\mu _{RPA}(\Gamma )$ is always positive, that is, the electrostatic
attraction, along with the specific binding energy $\varepsilon $, always
promotes crosslinking. From the numerical solution one obtains that $\mu
_{RPA}(\Gamma )$ is a monotonically decreasing function of $\Gamma $ for all
$\phi $, which is explained by the fact that as more ions associate they
contribute less to the long-range attraction (which can be seen directly
from $F_{RPA}$ in (\ref{frpa3})).

\section{Results and discussion}

\label{express}

Due to its complete symmetry, only macroscopic phase separation is possible
in the considered system. In our ternary incompressible system\ of polymer,
salt and solvent we have two independent components, which we choose to be
polymer and salt. When macroscopic phase separation (precipitation) occurs
the two coexisting phases differ in concentrations of both polymer and salt.
Thus, generally speaking, we have to calculate the phase diagrams of a
ternary incompressible system. However, we are mostly interested in two
aspects of the precipitation process: the influence of different competing
complexation mechanisms on the density of the formed precipitate (studied in
the next section) and determination of the conditions of solubility of the
complexes with addition of salt (section \ref{phd}). In both of these cases
we can make specific additional assumptions which allow us to investigate
the mentioned problems in a simple and clear manner.

\subsection{Density of precipitate: effect of crosslinking and van der Waals
attraction}

In our model we consider three complexation driving forces: long-range
electrostatic attraction between co-ions, strongly non-linear short-range
attraction leading to ion-crosslinking and van der Waals attraction between
all monomeric units. In this section, we look at the density of the formed
complex $\phi $, in particular how $\phi $ depends on the relative
importance of the three complexation factors as well as how the density is
influenced by the addition of salt. We can significantly simplify the
analysis if we make the following two assumptions (similar to the ones
employed in our previous work).\cite{mine} First, we assume an infinite
degree of polymerization $N=\infty $. Second, the total concentration of
polymer chains in the whole solution (system) is assumed to be small. Since
the entropy of the chains represents the only driving force for dissolution
of the polymer chains from the precipitate,\ the first assumption amounts to
assuming zero polymer concentration in the supernatant (which for a finite $N
$ would be a\ polymer-poor phase). The second assumption is equivalent to
assuming that the salt volume fraction in the supernatant is equal to the
salt volume fraction in the whole system, which we thus denote simply as $%
\phi _{s}$. Note that the salt volume fraction in the precipitate $\phi
_{s}^{(p)}$ can differ considerably from $\phi _{s}$, which, as we show
below, has a significant effect on $\phi $. Given our assumptions, $\phi $
and $\phi _{s}^{(p)}$ can be found by equating the pressure and the chemical
potential of the salt in the coexisting phases:%
\begin{eqnarray}
p(\phi  &=&0,\phi _{s})=p(\phi ,\phi _{s}^{(p)})  \label{pr} \\
\mu _{s}(\phi  &=&0,\phi _{s})=\mu _{s}(\phi ,\phi _{s}^{(p)})  \label{mu} \\
\mu _{s} &=&\frac{\partial {\cal F}}{\partial \phi _{s}} \\
p &=&-{\cal F}+\sum_{i=1}^{2}\phi _{i}\frac{\partial {\cal F}(\{\phi _{i}\})%
}{\partial \phi _{i}}
\end{eqnarray}%
For convenience here and in the following we use a dimensionless equilibrium
free energy density defined by ${\cal F}=F(\Gamma _{eq})/(kTV/v)$. The
equilibrium free energy $F(\Gamma _{eq})$ is obtained from the minimization
(described in section \ref{secmin}) of the total free energy, which
according to (\ref{f11}) is given by the sum of (\ref{fref2}) and (\ref%
{frpa3}). We solve equations (\ref{pr}--\ref{mu}) numerically to obtain $%
\phi $ and $\phi _{s}^{(p)}$, considering $\phi _{s}$ as a parameter. The
results are given in Figures 1--3.

In Figure 1 we assume no van der Waals attraction $\chi =0$ and plot results
for varying bonding energy $\varepsilon =E/kT$. (Experiments on
polyelectrolyte adsorption\cite{m01,g98} and multilayer formation\cite{m02}
are consistent with the value of binding energy $\varepsilon $\ varying
between $\varepsilon =3$ and $\varepsilon =7$.) We set the Bjerrum length $%
l=3$, and the fraction of charged monomers $f=0.1$. The dependence of $\phi $
on $l$ and $f$ for the case of complexation without crosslinking was
investigated in our previous work.\cite{mine} Similar to the results for
that case, $\phi $ increases with increasing $l$ and/or $f$ for all $\phi
_{s}$, $\varepsilon $ or $\chi $. With good precision one can obtain results
for other values of $f$ simply by linearly scaling $\phi $ with $f$.

Figure 1(a) shows the polymer volume fraction in the precipitate $\phi $ as
a function of the salt volume fraction in the system $\phi _{s}$. We see
that for all values of $\varepsilon $ the complex density monotonically
decreases with increasing salt concentration. This is of course due to
Debye-H\"{u}ckel screening by salt, which makes the electrostatic attraction
weaker (the term $F_{RPA}$ in the total free energy). Note that as $F_{RPA}$
becomes weaker the electrostatic binding energy $\mu $, given by (\ref{mg}),
also becomes smaller, thus both the long-range and short-range
electrostatics become less efficient in forming the complex. This is
illustrated in Figure 1(b), where we plot conversion $\Gamma $ for curves of
plot Figure 1(a). We see that conversion also monotonically drops when more
salt is added to the system, the behavior being similar for all values of $%
\varepsilon $. With the increase of $\varepsilon $, conversion $\Gamma $
increases for all salt concentrations. However this does not directly
translate into a denser complex as we can see from Figure 1(a). Indeed for $%
\phi _{s}\approx 0$ we see that $\phi $ depends non-monotonically on $%
\varepsilon $\thinspace , the feature which will be explored in detail in
Figure 3. For large salt concentrations the density $\phi $ is larger for
greater $\varepsilon $, which is obviously due to increased crosslinking of
chains. The presence of crosslinks manifests itself in a feeble dependence
of $\phi $ on $\phi _{s}$ for larger values of $\varepsilon $, which
indicates indissolubility by salt of complexes formed primarily by specific
short-range attractions. At the same time for small binding energies (such
as $\varepsilon =0$ and $\varepsilon =3$) in Figure 1(a) the density $\phi $
drops rather abruptly and becomes very small, which indicates dissolution of
the complex with addition of salt (this problem will be studied in the next
section). The difference between the salt concentration in the precipitate $%
\phi _{s}^{(p)}$and that in the supernatant $\phi _{s}$ (which , according
to our assumption, is equal to the salt concentration in the whole system)
is presented in Figure 1(c). We see that for small $\varepsilon $ the
precipitate is first enriched with salt, which (as we showed previously\cite%
{mine}) is due to correlational Debye-H\"{u}ckel attraction. For small $%
\varepsilon $ with increasing salt concentrations $\phi _{s}^{(p)}$ becomes
smaller than $\phi _{s}$ and then as the complex becomes very diluted for
large $\phi _{s}$ there is only a negligible difference. For large $%
\varepsilon $ we have $\phi _{s}^{(p)}<\phi _{s}$ for all salt
concentrations. (Note that we plot the difference $\phi _{s}^{(p)}-\phi _{s}$
in Figure 1(c), of course $\phi _{s}^{(p)}$ increases with $\phi _{s}$) The
depletion of salt was previously shown to be due to hardcore interactions.%
\cite{mine} This depletion turns out to have a considerable effect on $\phi $%
. In Figure 1(a) we plot with dashed lines the curves for $\varepsilon =5$
and $\varepsilon =7$ obtained from equation (\ref{pr}) with the assumption $%
\phi _{s}^{(p)}=\phi _{s}$. The effect can be seen to be especially
substantial for larger values of $\phi _{s}$.

In Figure 2 we illustrate the effect of the $\chi $-parameter for the case
when the effect of association is small ($\varepsilon =0$). The complex is
seen to become denser with increasing short-range attraction for all values
of $\phi _{s}$. For $\chi <0.5$ (good \ solvent) the volume fraction $\phi $
strongly decreases with increasing salt (which is an indication of
dissolution of the complex for finite $N$). However, for $\chi \geq 0.5$
(bad solvent condition) the density is seen to be negligibly dependent on $%
\phi _{s}$ for large $\phi _{s}$, with the complex being stable with respect
to addition of salt. Comparing Figures 1(a) and 2 we observe that behavior
of $\phi $ with increasing $\phi _{s}$ is qualitatively the same when the
complex is formed by crosslinking ($\varepsilon $) or van der Waals
interactions ($\chi $). Note that we analyze here only marginally bad
solvent conditions, for which we can disregard the possibility of necklace
formation.\cite{necklace}

The relative strength of the long-range electrostatic attraction vs.
crosslinking is demonstrated in Figure 3. We plot the precipitate density $%
\phi $ in the salt-free solution as a function of bond energy $\varepsilon $%
. Different curves correspond to varying $\chi $, which spans good to
marginal solvent conditions. Increasing $\varepsilon $ means that a greater
number of charges form crosslinks. Indeed from the numerical solution we
obtain that in all cases $\Gamma $ monotonically increases with growing $%
\varepsilon $. Thus increasing $\varepsilon $ physically means changing the
driving force of complexation from long-range charge correlation to
crosslinking attraction (numerically, at $\varepsilon =10$ the conversion $%
\Gamma \approx 1$). Interestingly, as we see from Figure 3, the density $%
\phi $ (although it exhibits a shallow minimum as a function of $\varepsilon
)$ is rather insensitive to the value of $\varepsilon $. Thus, ion-pairing
and long-range correlations lead to polyelectrolyte complexes of similar
density and the two mechanisms can be difficult to distinguish
experimentally for salt-free systems. Curves for different $\chi $ show
qualitatively the same behavior, with $\phi $ increasing as the solvent
worsens.

\subsection{Phase diagram: dissolution with addition of salt\label{phd}}

As can be seen from Figures. 1(a) and 2 for small $\varepsilon $ and/or $%
\chi $ the density $\phi $ becomes very small with addition of enough salt,
which is indicative of precipitate dissolution for a finite $N$. In the
previous section we assumed $N=\infty $, so the precipitate never dissolved.
In this section we relax this assumption and look at the phase coexistence.

Let us first investigate the effect of $N$ on the phase
coexistence. In order to simplify the presentation let us assume
that the salt volume fractions in the polymer-rich and
polymer-poor phases are the same. Thus we treat $\phi _{s}$ as a
parameter and obtain the coexisting polymer concentrations by
constructing a common tangent to the equilibrium free energy
density ${\cal F}$. An example of resulting phase diagrams for
varying $N$ is shown in Figure 4 (the values of all parameters are
given in the plot). The area below the coexistence line (for a
given $N$) corresponds to phase separation (complexation at low
salt concentrations), above --- to a homogeneous solution (the
precipitate dissolves at high salt $\phi _{s}$). Owing to our
simplifying assumption of equal $\phi _{s}$ in the two phases, the
tie lines giving the coexisting polymer concentrations\ are
parallel to the $\phi $-axis. We see that the dilute phase has
negligible polymer
concentration even for rather small $N$, which justifies the assumption $%
N=\infty $ of the previous section. When enough salt is added to the
solution the precipitate dissolves. As we can see from Figure 4 the salt
concentration needed to dissolve the precipitate strongly depends on $N$ and
(as Figures 1 and 2 show) it also depends on $\varepsilon $ and $\chi $.

In Figure 5 we determine the conditions on $\varepsilon $, $\chi $, and $N$
ensuring stability of the precipitate at a given concentration of salt. For
each curve (corresponding to a certain $\chi $) the precipitate exists at $%
\phi _{s}=0.1$ if the values of $\varepsilon $ and $\chi $ lie in the area
above the curve, and the precipitate is dissolved in the area below the
curve. (The value $\phi _{s}=0.1$ is taken as an example and it applies to
all curves). The increase of either of $\varepsilon $, $\chi $, or $N$
stabilizes the complex to addition of salt. We observe that\ the stability
of the precipitate is quite sensitive to the values of parameters in the
experimentally most relevant region $\varepsilon \approx 3$, $\chi \approx
0.5$, and $100<N<$ $1000$. The results of Figure 5 can be used for
experimental design of complexes stable to salt.

We obtained Figure 5 by considering the spinodal stability of our two
component system. The spinodal points are found from the following equation%
\begin{equation}
J(\phi ,\phi _{s})=\left\vert
\begin{array}{cc}
\frac{\partial ^{2}{\cal F}}{\partial \phi ^{2}} & \frac{\partial ^{2}{\cal F%
}}{\partial \phi \partial \phi _{s}} \\
\frac{\partial ^{2}{\cal F}}{\partial \phi \partial \phi _{s}} & \frac{%
\partial ^{2}{\cal F}}{\partial \phi _{s}^{2}}%
\end{array}%
\right\vert =0  \label{spin}
\end{equation}%
with ${\cal F}=F(\Gamma _{eq})v/(kTV)$ being the reduced equilibrium free
energy. In Figure 5 in the area above the curve for a certain $\chi $ the
equation $J(\phi ,\phi _{s}=0.1)=0$ has two solutions, while in the area
below the curve it has no solutions for physical values of $\phi $. It
should be noted that in our two component system this condition on the
existence of spinodal at a given salt is strictly speaking not equivalent to
existence of phase separation (to demand that the critical point be at $\phi
_{s}=0.1$ is yet another different condition). However, in the considered
system dissolution occurs at very low $\phi $,\ so the three conditions
yield numerically close results.

\section{Conclusions}

\label{conclusions}

Complexation in solutions of oppositely charged polyelectrolytes can be
accompanied by thermoreversible crosslinking of oppositely charged monomeric
ions. A local electrostatic binding energy can exist between oppositely
charged units when they tend to be dehydrated in the vicinity of each other
or due to non-classical specific interactions. In this work we have
investigated how the three different mechanisms (long-range electrostatics,
crosslinking and backbone hydrophobicity) define the properties of
polyelectrolyte complexes at different salt concentrations.

In our approach we obtain self-consistently the fraction of crosslinked
charged monomers (conversion) and the Debye-H\"{u}ckel collective
fluctuations contribution to the free energy (which depends on conversion).
Accordingly, the degree of conversion is determined both by the local
binding energy and\ by long-range electrostatics. We find that the
long-range charge fluctuations always promote crosslinking. Given that the
magnitude of the Debye-H\"{u}ckel contribution decreases with increasing
salt, the fraction of crosslinked monomers also monotonically decreases with
increasing salt.

The polymer concentration in the precipitate is largest at low salt
concentration, when the screening of interactions between monomeric ions is
weakest. The complex concentration generally decreases monotonically with
increasing salt concentration. The rate of complex dilution with addition of
salt and the concentration of monomers in the precipitate at high salt are
strongly dependent on the value of the van der Waals attractions and on the
binding energy. Non-selective net van der Waals attraction between the
monomers of both the positively and negatively charged chains enhances the
complexation in a way broadly similar to crosslinking due to local binding
energy. The dilution is very rapid and the monomer concentration of the
complex goes to zero in the case of zero binding energy and zero net van der
Waals attraction (good solvent condition). Instead, for large values of
either type of the short-range attractions, as the salt concentration
increases the monomer concentration in the complex generally nearly
saturates to a nonzero constant value (or slightly increases for
sufficiently large $\chi $). Our results are consistent with experimental
observations\cite{d97,d98,d02,d04} in which, depending on the type of
polymers used, with the addition of salt the complexes can either dissolve,
or their density can remain stable. In some cases initial dissolution and
subsequent re-entrant precipitation is observed\cite{d00}, the type of
behavior obtained in our theory for marginal solvent.

There is an important competition between complexation due to charge
fluctuations and non-linear thermoreversible linking, which is especially
interesting in good solvent conditions. Unassociated charged monomeric
groups induce complexation due to long-range electrostatics. However, once
they form crosslinks they practically do not contribute to long-range
attraction. Therefore at high conversion rates complexation is mostly due to
effective crosslinking attraction in an effectively neutral polymer
solution. This competition leads to an interesting non-monotonic behavior of
monomer concentration in the precipitate with increasing the non-linear
binding energy in the case of zero salt concentration: with the increasing
binding energy the monomer concentration in the precipitate passes through a
minimum. Remarkably, the variation in the complex density is rather small,
that is, crosslinking and long-range electrostatic attractions give rise to
complexes of similar density.

Another important effect in polyelectrolyte complexation is the difference
in salt concentration inside and outside the precipitate. When the
non-linear binding energy is small the difference in salt concentration in
and out of the precipitate is negligible. However, for large binding
energies (or large values of $\chi $) this difference rapidly grows as the
overall salt concentration increases, the precipitate being depleted of salt
due to increasing importance of hardcore interactions.\cite{mine} We find
that for large binding energies and/or in a bad solvent the difference
between salt concentrations in the complex and in the bulk has a significant
effect on the density of the complex at high salt concentrations. An
interesting limit to analyze includes the addition of non-linear
correlations among the ion pairs when the fraction of charged units
increases as in the case of strongly charged chains in oppositely charged
multivalent ion solution\cite{Solis00,Solis01} where even denser
precipitates are expected.

{\bf Acknowledgments}. This work was supported by NSF grant DMR 041446. A.
E. acknowledges partial financial support of NSF grant EECO 118025.

\appendix

\section{Calculation of the structural correlation functions}

\label{app}

In this section we calculate the structural correlation functions of the
solution of associating oppositely charged polyelectrolytes. Charged groups
on chains are treated as stickers that can associate with ionic groups of an
opposite sign. We consider only the symmetric case of oppositely charged
homopolymers of equal degree of polymerization $N=N_{1}=N_{2}$, present in
solution with equal concentrations $\rho _{1}=\rho _{2}=\rho /2$. (Indices $1
$ and $2$ refer to positively and negatively charged chains, respectively.)
We also assume that the chains have the same chemical structure, that is the
same bond length $b=b_{1}=b_{2}$, and the distance between charges $%
a=bf^{-1/2}$ ($f$ is the fraction of charged monomers on the chains). We
assume Gaussian statistics for chains.

The expression for the structural correlation function (\ref{gg}) between
two different types of monomers $\alpha $ and $\beta $ reads (for detailed
derivation see ref \CITE{we})%
\begin{eqnarray}
g_{\alpha \beta }(q) &=&\sum_{C}\rho (C)g_{\alpha \beta }^{C}(q)  \label{gab}
\\
g_{\alpha \beta }^{C}(q) &=&\sum_{i,j}\left\langle e^{i{\bf q}({\bf r}%
_{i}^{\alpha }-{\bf r}_{j}^{\beta })}\right\rangle _{C}  \label{gabc}
\end{eqnarray}%
Summation in (\ref{gab}) is over all topologically different clusters formed
due to association of polymers, with $\rho (C)$ being the number
concentration of a cluster having structure $C$ and $g_{\alpha \beta }^{C}(q)
$ the molecular structural correlation function. In (\ref{gabc}) the
summation runs over all monomers of types $\alpha $ and $\beta $ of the
cluster $C$. The average is over the conformations of the cluster, it can be
written as%
\begin{equation}
\left\langle e^{i{\bf q}({\bf r}_{i}^{\alpha }-{\bf r}_{j}^{\beta
})}\right\rangle _{C}=\frac{\int e^{i{\bf q}({\bf r}_{i}^{\alpha }-{\bf r}%
_{j}^{\beta })}f_{C}(\Gamma _{C})d\Gamma _{C}}{\int f_{C}(\Gamma
_{C})d\Gamma _{C}}=\frac{1}{V}\int e^{i{\bf q}({\bf r}_{i}^{\alpha }-{\bf r}%
_{j}^{\beta })}f_{C}(\Gamma _{C})d\Gamma _{C}
\end{equation}%
where $f_{C}(\Gamma _{C})$ is the probability function of finding the
cluster in conformation $\Gamma _{C}$, and the integration is over the
entire configurational space of the cluster $C$.

To calculate the correlation functions (\ref{gab}) we employ the grand
canonical diagrammatic technique described in ref \CITE{we}. (For details of
the diagrammatic technique as well as applications to other related systems
see refs \CITE{ee99,we,e04}.) As is shown in ref \CITE{we} the correlation
function can be expressed as the sum of all two-root diagrams%
\begin{equation}
g_{\alpha \beta }(q)=\sum_{n}z^{n}\sum_{C_{n}^{\alpha \beta }}\frac{%
W(C_{n}^{\alpha \beta })}{S(C_{n}^{\alpha \beta })}\left\langle e^{i{\bf q}(%
{\bf r}_{i}^{\alpha }-{\bf r}_{j}^{\beta })}\right\rangle _{C_{n}^{\alpha
\beta }}
\end{equation}%
where $z$ is the fugacity of the chain ($z=\exp (-\mu /kT)$, $\mu $ is the
chemical potential), $W(C_{n}^{\alpha \beta })$ is the statistical weight of
the cluster $C_{n}^{\alpha \beta }$ with two marked monomers of types $%
\alpha $ and $\beta $, and $S(C_{n}^{\alpha \beta })$ is its symmetry index.

In order to calculate $g_{\alpha \beta }(q)$ it is convenient to introduce
the following generating functions. Let us introduce the generating function
of all one-root diagrams $t$ (diagrams with one marked monomer), which can
be calculated recursively as%
\begin{equation}
t=1+\omega zt^{N-1}\frac{N}{2}  \label{tt}
\end{equation}%
Here $\omega $ is the statistical weight of the crosslink: $\omega =\exp
(\varepsilon )$, where $\varepsilon $ is the absolute value of the
dimensionless crosslink bond energy $\varepsilon =\left\vert E\right\vert
/kT $. Now we can write the sum of all labeled diagrams with two labels
belonging to the same chain as%
\begin{equation}
\Sigma _{g}(q)=zt^{N-2}\frac{1}{2}\sum_{i,j}\left\langle e^{i{\bf q}({\bf r}%
_{i}-{\bf r}_{j})}\right\rangle _{chain}=zt^{N-2}\frac{N}{2}g(q)  \label{sg}
\end{equation}%
where we have introduced the correlation function of one homopolymer chain $%
g(q)$. Note that the two labeled points are free, that is they have no
diagrams attached to them. We need to introduce also a closely related\ to $%
\Sigma _{g}$ sum of all diagrams where the two labels cannot belong to the
same monomer,%
\begin{eqnarray}
\Sigma _{h}(q) &=&zt^{N-2}\frac{1}{2}\sum_{i\neq j}\left\langle e^{i{\bf q}(%
{\bf r}_{i}-{\bf r}_{j})}\right\rangle _{chain}=zt^{N-2}\frac{N}{2}h(q) \\
h(q) &=&g(q)-1  \label{hf}
\end{eqnarray}

Since the system is symmetric we need to calculate only two functions $%
g_{11}=g_{22}$ and $g_{12}=g_{21}$. Using definitions (\ref{tt}--\ref{hf})
the autocorrelation function $g_{11}(q)$ can be written as the following
series%
\begin{equation}
g_{11}(q)=t^{2}\Sigma _{g}+t^{2}\Sigma _{g}(\omega ^{\prime }\Sigma
_{h})\omega ^{\prime }\Sigma _{g}+t^{2}\Sigma _{g}(\omega ^{\prime }\Sigma
_{h})^{3}\omega ^{\prime }\Sigma _{g}+...  \label{g111}
\end{equation}%
Here, the first term is the sum of all diagrams in which the two labels
belong to the same chain. Next terms result from summation of all diagrams
with labels belonging to different chains. In our model only oppositely
charged chains can associate with each other. The second term is the sum of
all diagrams in which the two labels (marking monomers on chains of type $1$%
) are separated by a chain of opposite charge (type $2$). The next term in (%
\ref{g111}) comes from summation of all diagrams with the insert between the
labeled chain comprised of three chains (sequence $2-1-2$). Higher terms
correspond to summation of diagrams with a higher number of chains in the
insert between the labeled chains.

In (\ref{g111}) we introduced the effective statistical weight of the
crosslink%
\begin{equation}
\omega ^{\prime }=\omega e^{-q^{2}b^{2}/6}
\end{equation}%
which takes into account the correlations of monomers in a crosslink, which
we assume to be Gaussian ($b$ is the bond length). Note that since one
monomer can form only one crosslink the presence in a diagram of a chain
connecting the two labeled chains corresponds in eq \ref{g111} to a
generating function $\Sigma _{h}\,$, in which summation runs over$\ i\neq j$%
. It is easy to see that the term in (\ref{g111}) corresponding to the sum
of all diagrams separated by $2n-1$ chains reads $t^{2}\Sigma _{g}(\omega
^{\prime }\Sigma _{h})^{2n-1}\omega ^{\prime }\Sigma _{g}$. It is easy to
sum the infinite series (\ref{g111}) as%
\begin{equation}
g_{11}(q)=t^{2}\Sigma _{g}\frac{1+\omega ^{\prime }\Sigma _{g}\omega
^{\prime }(\Sigma _{g}-\Sigma _{h})}{1-\left( \omega ^{\prime }\Sigma
_{h}\right) ^{2}}  \label{g11}
\end{equation}%
Analogously for $g_{12}(q)$ we obtain%
\begin{equation}
g_{12}(q)=t^{2}\Sigma _{g}\omega ^{\prime }\Sigma _{g}+t^{2}\Sigma
_{g}(\omega ^{\prime }\Sigma _{h})^{2}\omega ^{\prime }\Sigma
_{g}+t^{2}\Sigma _{g}(\omega ^{\prime }\Sigma _{h})^{4}\omega ^{\prime
}\Sigma _{g}+...  \label{g121}
\end{equation}%
where the first term corresponds to all diagram with the labels belonging to
two neighboring chains (of opposite charge), the next term --- all diagrams
with two chains separating the labeled chains and so on. The series (\ref%
{g121}) can be summed as
\begin{equation}
g_{12}(q)=t^{2}\Sigma _{g}\frac{\omega ^{\prime }\Sigma _{g}}{1-\left(
\omega ^{\prime }\Sigma _{h}\right) ^{2}}  \label{g12}
\end{equation}%
To use the correlators (\ref{g11}) and (\ref{g12}) in the free energy we
need to change from the variables $z$ and $t$ to the number concentration of
charged monomers $\rho _{1}$ (for our case $\rho _{1}=\rho _{2}$) and
conversion $\Gamma $. Conversion $\Gamma $ is defined as the fraction of
charged monomers in crosslinks%
\begin{equation}
\Gamma =\frac{\rho _{1}}{\rho ^{(2)}}=\frac{\rho _{2}}{\rho ^{(2)}}
\label{dg}
\end{equation}%
where $\rho ^{(2)}$ is the number of crosslinks. As is shown in refs %
\CITE{ee99,we} the concentrations are given by%
\begin{eqnarray}
\rho _{1} &=&\frac{zt^{N}}{2}N  \label{r1} \\
\rho ^{(2)} &=&\omega \left( \frac{1}{2}zNt^{N-1}\right) ^{2}  \label{r2}
\end{eqnarray}%
Combining it with (\ref{tt}) and (\ref{sg}) we obtain%
\begin{eqnarray}
t^{2}\Sigma _{g} &=&\rho _{1}g(q) \\
\omega ^{\prime }\Sigma _{h} &=&\Gamma e^{-q^{2}b^{2}/6}h(k)=\Gamma ^{\prime
}h(k)  \label{ec}
\end{eqnarray}%
which substituted into (\ref{g11}) and (\ref{g12}) yields%
\begin{eqnarray}
g_{11}(q) &=&\rho _{1}g(q)\frac{1+\left( \Gamma ^{\prime }\right) ^{2}h(q)}{%
1-\left[ \Gamma ^{\prime }h(q)\right] ^{2}}  \label{gf11} \\
g_{12}(q) &=&\rho _{1}g(q)\frac{\Gamma ^{\prime }g(q)}{1-\left[ \Gamma
^{\prime }h(q)\right] ^{2}}  \label{gf12}
\end{eqnarray}%
Note that in (\ref{ec}) we introduced the effective conversion $\Gamma
^{\prime }$, which takes into account the correlations of ions in the
crosslink. The divergence of the correlators at $\Gamma ^{\prime }h(q=0)=1$,
i.e. $\Gamma (N-1)=1$ corresponds to gelation.\cite{ee99,we,eprl}

The value of $\Gamma $ can be obtained from the definition of $\Gamma $
(given by eq \ref{dg}) using relations (\ref{tt}), (\ref{r1}), and (\ref{r2}%
). Conversion turns out to be determined by the unmodified mass action law%
\begin{equation}
\frac{\Gamma }{(1-\Gamma )^{2}}=\rho _{1}\omega =\rho _{1}e^{\varepsilon }
\label{malo}
\end{equation}%
This reflects the fact that in this Appendix we considered \ an ideal
associating system (no other interactions except association are present).
Conversion for the solution of associating polyelectrolytes is obtained from
the minimization of the total free energy, which includes interaction terms.
The resulting modified mass action law (\ref{mal}) differs from eq \ref{malo}
in that it \ has a long-range electrostatic contribution to the effective
binding energy.

\newpage

{\LARGE Figure Captions}

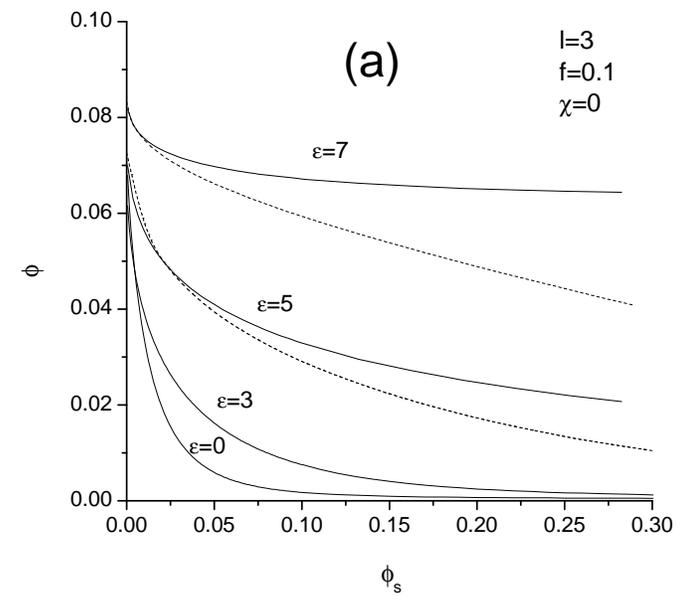
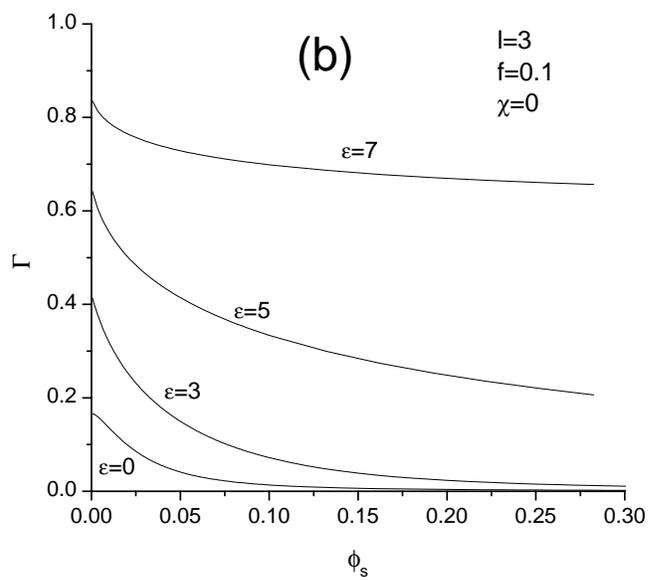
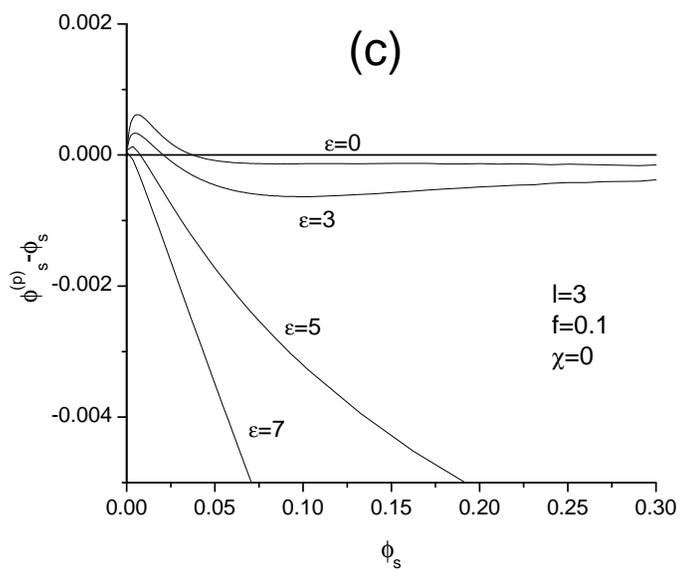
{\bf Figure 1.} (a) Polymer volume fraction in the precipitate $\phi $ as a
function of the salt volume fraction in the supernatant $\phi _{s}$ for
different binding energies $\varepsilon $. Dash lines correspond to the
assumption of equality of salt concentrations in the precipitate and
supernatant: $\phi _{s}^{(p)}=\phi _{s}$. (b) Conversion $\Gamma $ for
curves of plot (a). (c) The difference of salt volume fractions in the
precipitate and supernatant $\phi _{s}^{(p)}-\phi _{s}$ for the curves of
plot (a).

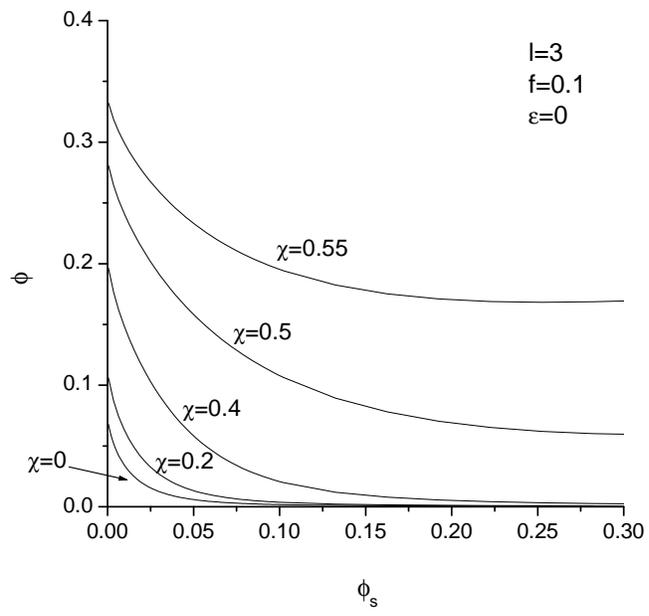
{\bf Figure 2.} Effect of the Flory-Huggins $\chi $-parameter on the change
of polymer volume fraction in the precipitate $\phi $ with increasing salt
in the system $\phi _{s}$.

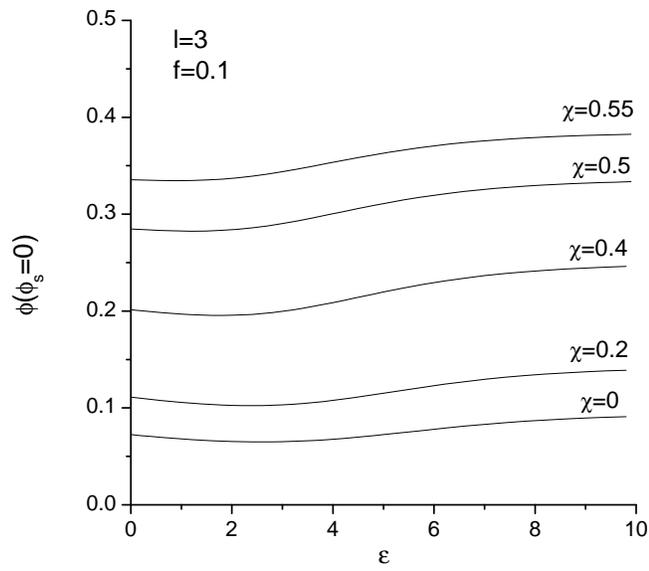
{\bf Figure 3.} Variation of the density of the precipitate for salt-free
system with changing binding energy $\varepsilon $. Different curves
correspond to different values of the $\chi $-parameter.

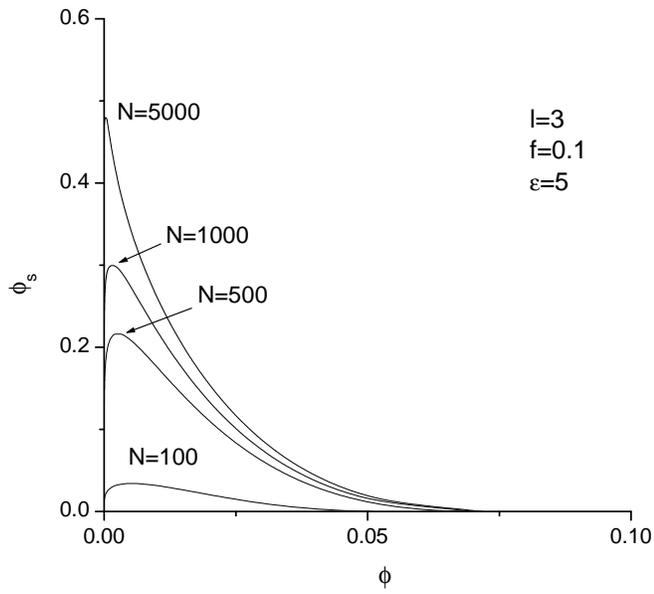
{\bf Figure 4.} Coexistence lines for phases with different polymer volume
fractions $\phi $ at a given salt volume fraction $\phi _{s}$. The effect of
varying chain length $N$ is illustrated.

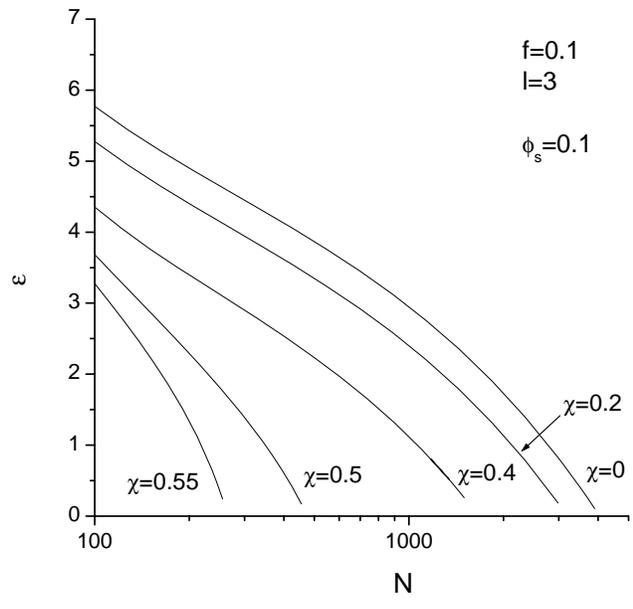
{\bf Figure 5.} Stability of the precipitate to addition of salt. For all
values of the bond energy $\varepsilon $ and the chain length $N$ \ below
the curve for a corresponding $\chi $-value the precipitate dissolves when
the\ salt concentration is increased beyond $\phi _{s}>0.1$.

\end{document}